\documentclass[10pts,aps,twocolumn,noshowpacs,noshowkeys,superscriptaddress
]{revtex4}
\usepackage{amsfonts}
\usepackage{amsmath}
\usepackage{amssymb}
\usepackage{graphicx}

\usepackage{textcomp}%
\newcommand{\vv}[1]{\mathbf{#1}}

\begin{document}
\def\neel{Institut N\'{e}el,  Universit\'{e} Grenoble Alpes - CNRS:UPR2940, 38042 Grenoble, France}
\def\ilm{Institut Lumi\`{e}re Mati\`{e}re,  UMR5306, CNRS - Universit\'{e} Claude Bernard Lyon 1, 69622 Villeurbanne, France}
\author{A.~Gloppe}
\affiliation{\neel}
\author{P. Verlot}
\affiliation{\neel}
\author{E. Dupont-Ferrier}
\affiliation{\neel}
\author{A. Siria}
\affiliation{\ilm}
\author{P. Poncharal}
\affiliation{\ilm}
\author{G. Bachelier}
\affiliation{\neel}
\author{P. Vincent}
\affiliation{\ilm}
\author{O. Arcizet}
\affiliation{\neel}
\email{olivier.arcizet@neel.cnrs.fr}
\title{Bidimensional nano-optomechanics and topological backaction in a non-conservative radiation force field}

\begin{abstract}
{Optomechanics, which explores the fundamental coupling between light and mechanical motion, has made important advances in both exploring and manipulating macroscopic mechanical oscillators down to the quantum level. However, dynamical effects related to the vectorial nature of the optomechanical interaction remain to be investigated. Here we study a nanowire with sub-wavelength dimensions strongly coupled to a tightly focused beam of light, enabling ultrasensitive readout of the nanoresonators dynamics. We experimentally determine the vectorial structure of the optomechanical interaction and demonstrate that bidimensional dynamical backaction governs the nanowire dynamics. Moreover, the non-conservative topology of the optomechanical interaction is responsible for a novel canonical signature of strong coupling between the nanoresonator mechanical modes, leading to a topological instability. These results have a universal character and illustrate the increased sensitivity of nanomechanical devices towards topologically varying interactions, opening fundamental perspectives in nanomechanics, optomechanics, ultrasensitive scanning probe force microscopy and nano-optics.}
\end{abstract}
\maketitle

\newsavebox{\smlmat}
\savebox{\smlmat}{$\left(\begin{smallmatrix}0&1\\-1&0 \end{smallmatrix}\right)$}

The optomechanical interaction  offers a canonical support for understanding fundamental concepts of modern physics: the general picture of a movable object interacting with light has early been recognized for formalizing and understanding foundations of quantum mechanics \cite{bohr1949discussion} or quantum limits in continuous measurements \cite{caves1982quantum,braginsky1995quantum,jaekel2007quantum}. While the possibility to manipulate atoms or micro-particles with light has been soon realized and implemented, the faintness of the optomechanical interaction with macroscopic objects has long retained the pioneering proposals far from experimental reach.
In recent years, the avenue of cavity optomechanics has successfully addressed this experimental challenge by coupling the mechanical oscillator to a high-finesse optical cavity, which serves as a noise-free optical amplifier enabling quantum limited optomechanical interaction.
Since their emergence in the late 90's \cite{dorsel1983optical,Cohadon1999}, these systems have been continuously improved and diversified \cite{Kippenberg2008}, eventually giving rise to laser cooling of macroscopic oscillators \cite{Arcizet2006,Gigan2006,Thomson2008} close to their quantum ground state \cite{Teufel2011,chan2011laser,Verhagen2012} or classical and quantum optomechanical operation \cite{verlot2009,marino2010classical,purdy2013observation,verlot2010backaction,
Weis2010,safavi2011electromagnetically}.\\
To optimize the optomechanical interaction, these experiments have been developed along the intuitive criterion of maximizing the overlap between the mechanical deformation and one single mode of the electromagnetic field \cite{Pinard1999}, the cavity mode. By doing so, this inherently scalar approach has left aside one of the most intriguing property of the radiation force field: its vectorial character and the non-conservative nature of the optomechanical interaction which only appear in dimensions larger than one.
However this specificity, arising from the Lorentz force contribution to the light-matter interaction \cite{Ashkin1986,Cohen-Tannoudji2004,Novotny2006}, has fundamental implication for the field of optomechanics.  For example, the path-dependence of the accumulated work strongly affects the thermodynamic state of systems evolving in non-conservative force fields \cite{Roichman2008,Sun2009}, which can even alter their dynamical stability \cite{Krechetnikov2007}.
These deviations are expected to be enhanced in the emerging field of nano-optomechanics, which permits exploring the light-matter interaction on dimensions smaller than the optical wavelength, but none of the existing experiments \cite{Li2008,Anetsberger2009,Favero2009, Gavartin2011,Ramos2012} have so far explored the vectorial and non-conservative character of the optomechanical interaction and their fundamental implications on the measurement backaction.\\
\begin{figure*}[t!]
\begin{center}
\includegraphics[width=0.92 \linewidth]{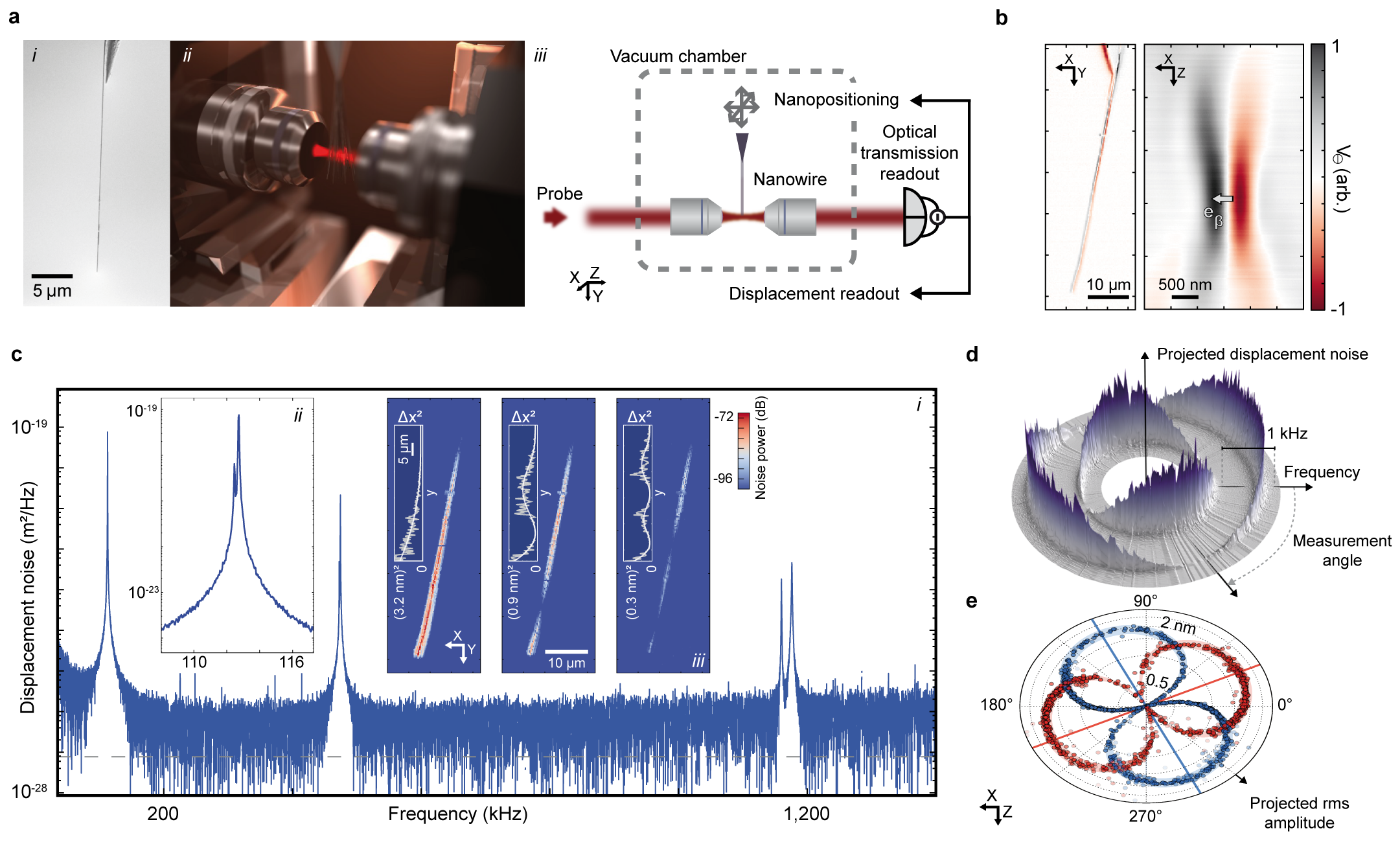}
\caption{ \textbf{An ultrasensitive optical readout of motion at the nanoscale.}  (a) Our nanomechanical oscillators consist of low-dissipation suspended silicon carbide (SiC) nanowires with lengths and diameters in the $10-100\,\rm \mu m / 50-300\,\rm nm $ range. Their extremity is positioned with a computer-controlled piezostage within the waist of a strongly focused laser beam ($w_{0}\approx 550\,\mathrm{nm}$ at 633\,nm) generated with a high numerical aperture microscope objective ($0.75\,\rm NA$). The transmitted light is collected through a second objective and detected on a differential quadrant photodetector (QPD). The nanowire position is controlled with an accuracy below $10\,\mathrm{nm}$ and position tracking routines ensure long term stability. The ensemble is mounted in a vacuum chamber to reduce air damping.  (b) Recording the DC differential transmission $V_\ominus$ while scanning the nanowire position in the vertical (left) and horizontal (right) planes allows imaging the nanowire.  The vector represents the transmission gradient $\vv{\nabla} V_\ominus$ at the center of the waist where the noise spectrum of the QPD AC output port is acquired (panel (c), 36\,Hz resolution bandwidth) revealing the Brownian motion of the first 3 eigenmode families with a very large signal-to-noise. Recording the displacement noise for each mode in the vertical plane (insets iii) permits determining their longitudinal vibration profiles (see SI).  Each mode family is composed of two peaks (inset ii), corresponding to the two eigenmodes vibrating along perpendicular directions. (d) Projected displacement noise spectra $S_{\delta r_\beta}[\Omega]$ reported as a function of the direction $\vv{e_\beta}$ of the measurement vector for a constant injected power of $100\,\rm \mu W$. 900 positions are sampled in the horizontal plane, over which $\vv{e_\beta}$ samples any possible orientation while the local intensity varies by more than 4 orders of magnitude. Panel (e) shows the corresponding projected r.m.s. vibration amplitude for both polarizations also reported as a function of $\vv{e_\beta}$, fitted using $\Delta x_i |\vv{e_i}\cdot\vv{e_\beta}|$. This allows determining the eigenmodes orientations $\vv{e_{1,2}}$ and verifying their perpendicularity at the percent level. The effective masses deduced ($376\,\rm fg$ here for a $25\times 0.15\,\rm\mu m$ nanowire) are found to be position independent to within a $20\%$ error margin, underlying the absence of static heating due to light absorption.
\label{Fig1}}
\end{center}
\end{figure*}
In this article we present a versatile optomechanical approach that goes beyond the standard scalar description of cavity optomechanics. Its principle relies on coupling a suspended sub-wavelength sized crystalline nanowire to a strongly focused laser beam. This cavity-free and optically broadband approach gives rise to large optomechanical coupling in both axial and transverse vibration directions, enabling ultrasensitive readout of the nanomechanical motion for incident optical power in the $100\,\rm \mu W$ range only. Exploiting the uniquely low force noise of SiC nanowires \cite{Perisanu2007}, we operate our nanoresonator as an ultrasensitive vectorial force sensor which enables full topological reconstruction of the nano-optomechanical interaction within the beam waist area. We show that topological variations of the optomechanical interaction are responsible for a novel vectorial backaction mechanism which can not be described within the scalar frame of cavity optomechanics, representing a new canonical signature of strong coupling between oscillators in dimensions larger than one.
We demonstrate that force fields presenting a large rotational character can activate a bifurcation of the nanoresonator dynamics, illustrating the connections existing between dissipation, dimensionality and conservativity \cite{Krechetnikov2007}.\\
On the fundamental side, this work establishes nano-optomechanical experiments as a method of choice for  exploring  light-matter interaction at the nanoscale \cite{ren1996prediction,dogariu2012optically}. With ultrahigh vectorial force sensitivity in the few aN-range ($10^{-18}\,\rm N$) at room temperature, our system appears as particularly suited for investigating weak vectorial force fields, in particular in the emerging field of hybrid nanomechanics \cite{Wilson-Rae2004,rabl2010quantum,Arcizet2011,Yeo2013}. On a more applied side, this work shows that ultralow noise optical readout can be standardly transferred to NEMS-based detection schemes  \cite{Gavartin2012,Chaste2012,Ramos2012,hanay2012single}.\\

\begin{figure*}
\begin{center}
\includegraphics[width=0.9 \linewidth]{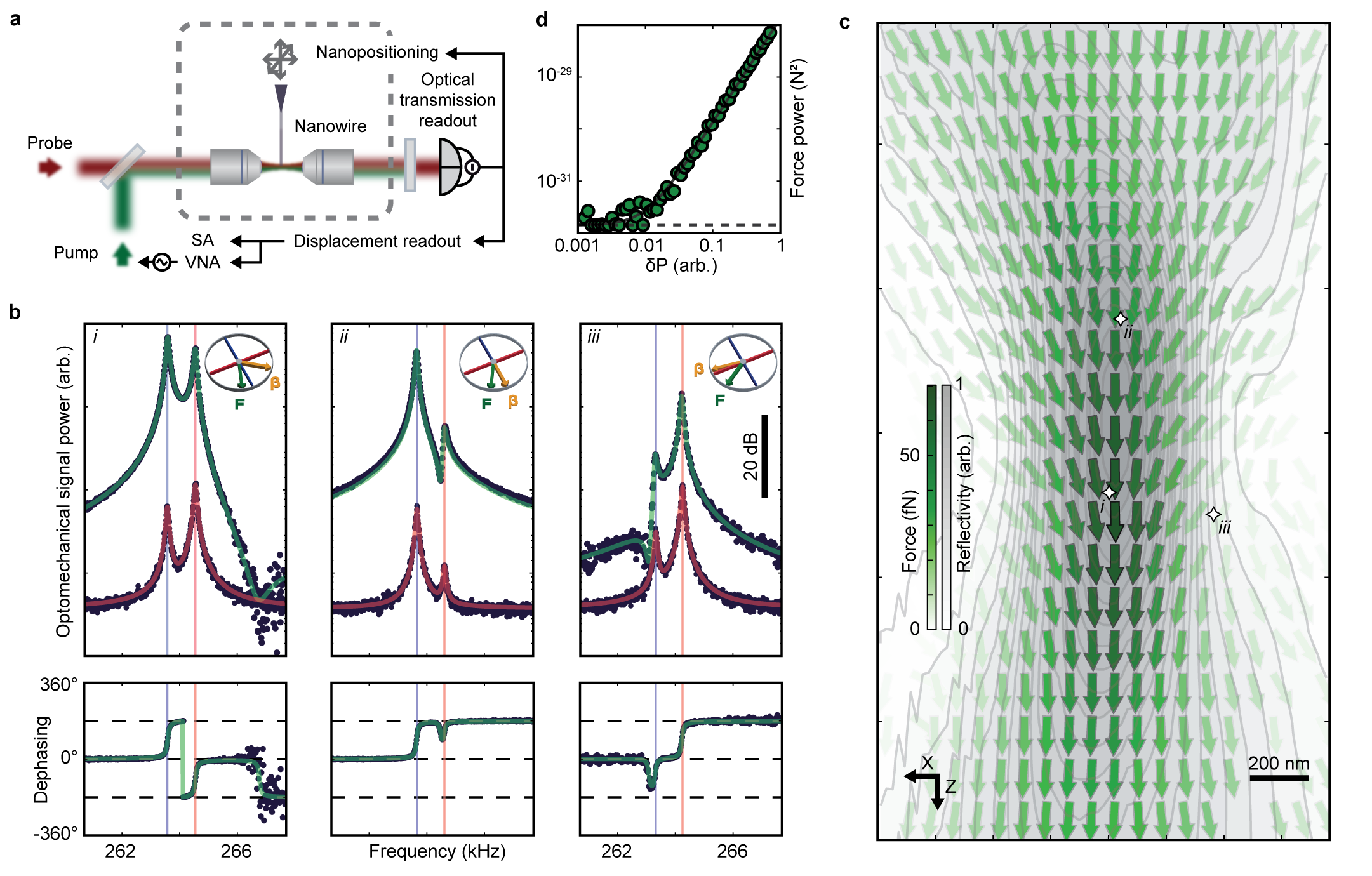}
\caption{  \textbf{Measurement of the bidimensional topology of the optomechanical interaction force field.} (a) Schematics of the experiment: a second laser (532\,nm), intensity modulated by means of an acousto-optic modulator is focused on the nanowire, whose motion is simultaneously recorded with a 633\,nm probe laser. (b) For each measurement point (i to iii, reported in panel c), Brownian motion (red fitted curve) and mechanical response (amplitude and phase) measurements (green fitted curves) are recorded on spectrum and network analyzers respectively. The fit (see text) allows inferring the local amplitude, direction and phase delay of the optical force exerted by the pump laser.
Arrows in the sketches indicate the deduced orientations for the local measurement $\boldsymbol{\beta}$ and force  $\vv{F}$ vectors. (c) This measurement sequence was reproduced on a 40x40 grid in a  $1.6\times 3\,\rm \mu m^2$ horizontal area around the pump laser waist (22 hours duration), in order to establish the spatial map of the optomechanical interaction between the 532\,nm focussed laser beam and the nanoresonator (data are averaged on a 2x2 grid). The normalized pump beam reflectivity map is also shown for comparison. (d) The linearity of the experiment was verified by varying the modulation strength and recording the force magnitude down to the thermal noise limit ($38\,\rm aN/\sqrt{Hz}$ here).
\label{Fig2} }
\end{center}
\end{figure*}

\textit{An ultrasensitive vectorial displacement sensor.---}
Our nano-optomechanical scheme  (Fig.\,1a) consists in positioning the extremity of a suspended silicon carbide nanowire at the waist of a  strongly focused laser beam  obtained with a high numerical aperture microscope objective while collecting the transmitted light \cite{nichol2012viewpoint,Arcizet2011,Raizen2011,Gieseler2012}. The wide bandgap of SiC allows the nanowire to sustain significant optical powers ($\lesssim\rm mW$) in the visible domain without incurring any damage.
Nanomechanical motion is measured using a differential transmission measurement technique \cite{treps2003quantum} based on a high gain dual quadrant photodiode. Its differential output voltage $V_\ominus$ depends  both on the piezo-controlled nanowire rest position $\mathbf{r_0}$ and on its bidimensional deflection $\vv{\delta r}$. For small vibration amplitudes, it can be expanded as $V_\ominus(\mathbf{r_0+\delta r}(t))\approx V_\ominus(\mathbf{r_0})+\boldsymbol{\beta}\cdot\mathbf{\delta r}(t)$.  The projective measurement vector $\boldsymbol{\beta}\equiv\left.\nabla V_\ominus\right|_\mathbf{r_0}\equiv \beta {\vv{e_\beta}}$ thus converts the bidimensional vibrations $\delta\vv{r}$ of the nanowire extremity into a measurable scalar quantity $\delta r_\beta\equiv\delta \vv{r}\cdot\vv{e_\beta}$.  Typical static differential transmission maps  $V_\ominus(\mathbf{r_0})$ acquired while scanning the nanowire rest position $\mathbf{r_0}$ in the vertical $(\mathbf{e_x},\mathbf{e_y})$ and horizontal $(\mathbf{e_x},\mathbf{e_z})$ planes are shown on Fig.\,1b (left and right respectively). Despite the sub-wavelength diameter of the nanowire, a large transmission contrast can be achieved, leading to a quasi suppression of the light transmission when positioned on the optical axis. As a consequence, the Brownian motion of the nanowire can be monitored with a very high sensitivity by measuring the spectrum of the differential photocurrent fluctuations $\delta V_\ominus(t)$, $S_{\mathrm{V}_\ominus}[\Omega]\equiv\int_{-\infty}^{+\infty}\mathrm{d}\tau e^{-i\Omega\tau}\langle \delta V_\ominus(t)\delta V_\ominus(t+\tau)\rangle$.
Fig.\,1c shows typical calibrated displacement spectrum acquired with the nanowire extremity  being centered in the optical waist. Sharp peaks are observed at  113, 473 and 1160\,kHz,  completing a series matching the eigenfrequencies expected for a $52\,\rm\mu\mathrm{m}$ long, $150\,\rm\mathrm{nm}$ diameter, singly-clamped SiC nanowire.  Moving the nanowire in the vertical plane allows to determine the spatial profiles of the first eigenmodes Brownian motion  (Fig.\,1c.iii) which serves for effective masses derivation (see SI).
Even for a  modest quality factors of $Q=2890$, the fluctuations of the fundamental vibrational mode are resolved with an extremely high signal-to-noise ratio of $72\,\mathrm{dB}$ with respect to the shot noise limited detection background. Recording the displacement noise at various incident optical powers ($10-100\,\rm\mu\mathrm{W}$) and environment pressure ($0.01-10\,\mathrm{mBar}$), we determine photothermal heating effects to be negligible and checked that our system is driven by an incoherent force which verifies the fluctuation-dissipation theorem (see SI), establishing thermal noise as the dominant external excitation.\\
Zooming over each peak reveals the systematic presence of split resonance doublets (Fig.\,1c), corresponding to the two mechanical polarizations of each longitudinal eigenmode family (frequencies $\Omega_{1,2}$, directions $\mathbf{e_{1,2}}$), whose degeneracy is lifted by the asymmetries in nanowire geometry and clamping conditions \cite{nichol2008displacement}.
The generic dynamics of the nanowire extremity vectorial deflection  $\mathbf{\delta r}$, restricted to the first eigenmode family (around $113\,\rm kHz$) can be described by:
$$\mathbf{\delta\ddot r}= -\vv{\Omega^2}\mathbf{\delta r}-\Gamma \mathbf{\delta\dot r}+{\mathbf{\delta F_{\rm th}}(t)}/{M}+{\vv{F}(\mathbf{r_0+\delta r},t)}/{M}
$$
where $M$ is the effective mass of the nanoresonator, $\vv{\Omega^2}\equiv\left(\begin{smallmatrix}\Omega_1^2&0\\0&\Omega_2^2\\ \end{smallmatrix}\right) $ the restoring force matrix in the  $\mathbf{e_{1,2}}$ base and $\mathbf{\delta F_{\rm th}}=\left( \delta F_{\rm  th}^1, \delta F_{\rm  th}^2 \right)$ the vector of Langevin forces driving each modes independently \cite{Kubo1966,Pinard1999} with a white spectral density of $S_{F_i}^{\rm th}=2 M \Gamma k_B T$ ($k_B$ is the Boltzmann's constant, $T=300\,\mathrm{K}$). For simplicity we have used here equal damping rates $\Gamma$. The effective masses of both polarizations are identical as confirmed experimentally (see SI). The term $\vv{F}(\mathbf{r_0+\delta r},t)$ represents a generic external force field in which evolves the nanoresonator,  which can vary in space and time.\\
At low light intensities optical forces can be neglected and the spectrum of the nanowire vibrations $\delta r_\beta$ projected along the direction $\vv{e_\beta}$ of the measurement vector, can be approximated by:
$$S_{\delta r_\beta}[\Omega]=S_{V_\ominus}[\Omega]/\beta^2=\sum_{i=1,2}{(\mathbf{e_i}\cdot\mathbf{e_\beta})^2\, \left|\chi_i[\Omega]\right|^2S_{F_i}^{\rm th}} $$
with $\chi_i[\Omega]\equiv1/\left( M\left(\Omega_i^2-\Omega^2-i \Omega \Gamma \right)\right)$ being the mechanical susceptibilities. The displacement sensitivities on each eigenmode are therefore related to their orientation with respect to the measurement vector $\boldsymbol{\beta}$ which is experimentally measured prior to any dynamical measurement (see SI).  It can be varied by moving the nanowire within the waist area, enabling a full reconstruction of the eigenmodes orientations and effective masses (see Fig.\,1d and SI), which are extremely important for any quantitative vectorial applications, such as force field imaging.

\textit{An ultrasensitive vectorial force sensor.---}
Measuring the nanowire deformations in two dimensions can naturally be used to measure vectorial forces, with sensitivities in the $\rm aN/\sqrt{Hz}$ range at room temperature. In this context, it is of primary importance to investigate and control the backaction process resulting from the force field associated with the vectorial measurement. In our case, it corresponds to the optical force field experienced by the nanowire, $\vv{F}(\vv{r_0})$ for an incident power $P_0$.
The laser beam under investigation is intensity-modulated by means of an acousto-optic modulator (see Fig.\,2a) such that $P(t)=P_0+\delta P\cos \Omega t$, with an amplitude $\delta P$  resulting in a modulated optical force field of amplitude $\delta\vv{F}(\vv{r_0})= \vv{F}(\vv{r_0})\,\delta P/ P_0$. The measurements were taken at low light intensities such that the optical force gradients do not perturb the nanowire stability (see below). The modulated force is frequency swept across the  first mechanical doublet by means of a network analyzer  (see Fig.\,2b) which simultaneously demodulates the nano-optomechanical signal:
\begin{equation}
\delta r_\beta[\Omega]=\sum_{i=1,2}{\chi_i[\Omega](\mathbf{\delta F}(\vv{r_0})\cdot\mathbf{e_i}+\delta F_{\rm th}^i[\Omega])\,  (\mathbf{e_i}\cdot\mathbf{e_\beta})}
\label{eq.response}
\end{equation}
where the detection background has been neglected.
Fitting this complex nano-optomechanical response, represented in Fig.\,2b(i-iii), yields to a complete determination of the local force field  $\vv{F}(\vv{r_0})$: its magnitude, its orientation and the possible delay between the intensity modulation and the optical force. Our measurement confirms that the optical force is in phase with the intensity modulation, as expected for scattering forces which respond instantaneously on the timescale of the nanowire dynamics. The linearity of the response has been verified by varying the modulation depth $\delta P$ over 3 orders of magnitude while measuring the induced force magnitude, see Fig.\,2d. Iterating this measurement over a $1.6\times 3.0\,\rm\mu m^2$ area around the laser waist (see Fig.\,2d) allows to determine the topology of the optomechanical interaction $\mathbf{F(r_0)}$, measured here at 532\,nm and $P_0=96\,\rm \mu W$. The force field presents a strongly converging/diverging vector flow in the upstream/downstream regions, respectively, reaching a maximum value of $70\,\rm fN$ at the waist, where it is aligned along the laser propagation axis. These measurements were also performed at 633\,nm (see SI), wavelength at which the optomechanical force field presents a similar pattern for our $150\,\rm nm$-diameter nanowire, with a maximum value of $14\,\rm fN$ for the same incident power.
It is interesting to note that in contrast to scalar 1D optomechanics, there exists locations for which the backaction force can be chosen perpendicular to an eigenmode direction without compromising the respective displacement readout sensitivity. This enables suppressing the classical  measurement backaction \cite{caniard2007observation}, which is critical in ultrasensitive force sensing experiments.

\textit{Dynamical backaction in a bidimensional optical force field. --- }
At large optical powers, the dynamical backaction of the optical force field onto the nanoresonator has to be taken into account. The optical force follows instantaneously any intensity change so that when undergoing Brownian motion in the optical waist, the nanowire experiences a position-dependent optical force:  $\mathbf{F}(\mathbf{r_0}+\mathbf{\delta r}) \approx  \vv{F}(\vv{r_0})+ \left. (\vv{\delta r}\cdot\nabla)\vv{F}\right|_\vv{r_0}$. The first term generates a static deflection of the nanowire extremity, that may potentially lead to static vectorial multistability \cite{dorsel1983optical}, which can be neglected in our case ($\chi_i[0] F_i)\lesssim 1{\AA}$). The second term represents the generalized dynamical vectorial backaction. The restoring force matrix is thus modified to:
\begin{equation}
\vv{\Omega^2}=\left(
 \begin{matrix}
 \Omega_{1}^2 -g_{11}& -g_{21}\\
 -g_{12} &  \Omega_{2}^2-g_{22} \\
 \end{matrix} \right),
\label{eq.matrix}
\end{equation}
where $g_{ij}\equiv M^{-1}  \left.\partial_i F_j \right|_\mathbf{r_0}$ are proportional to the injected light intensity. The spatial derivatives of the force field can be directly calculated from the vectorial map shown in Fig.\,2d, they are shown in Fig.\,3.
In analogy with the optical spring effect of cavity optomechanics  which originates from the radiation pressure gradient that exists in a detuned Fabry-Perot cavity \cite{sheard2004observation,Kippenberg2008,Arcizet2006}, diagonal terms $g_{ii}$ modify the effective spring constants as well known and profusely used in ultrasensitive force microscopy.
The non-diagonal terms, equal in the case of a conservative force field, ($\vv{rot}(\vv{F})_y\propto g_{12}-g_{21}=0$), are coupling both mechanical polarizations. These shear components vanish on the optical axis and present extrema on each sides of the waist, where the force field presents large vorticity.
In orders of magnitude, the optical force gradients reach a level of $\simeq 300\,\rm nN/m$ for $P_0=100\,\rm\mu W$. Although this number remains small compared to the bare oscillator stiffness $\sim 400\,\rm \mu N/m$, the backaction force noise experienced by the nanowire can reach $ \nabla\mathbf{F}\cdot\vv{\delta r}_{\mathrm{th}}[\Omega_{i}]\simeq 30\,\mathrm{aN}/\sqrt{\mathrm{Hz}}$, which is comparable to the magnitude of the Langevin force and is hence of utter importance for a number of ultrasensitive force microscopy experiments \cite{Rugar2004,nichol2012viewpoint,Arcizet2011}.

\begin{figure}[t!]
\begin{center}
\includegraphics[width= \linewidth]{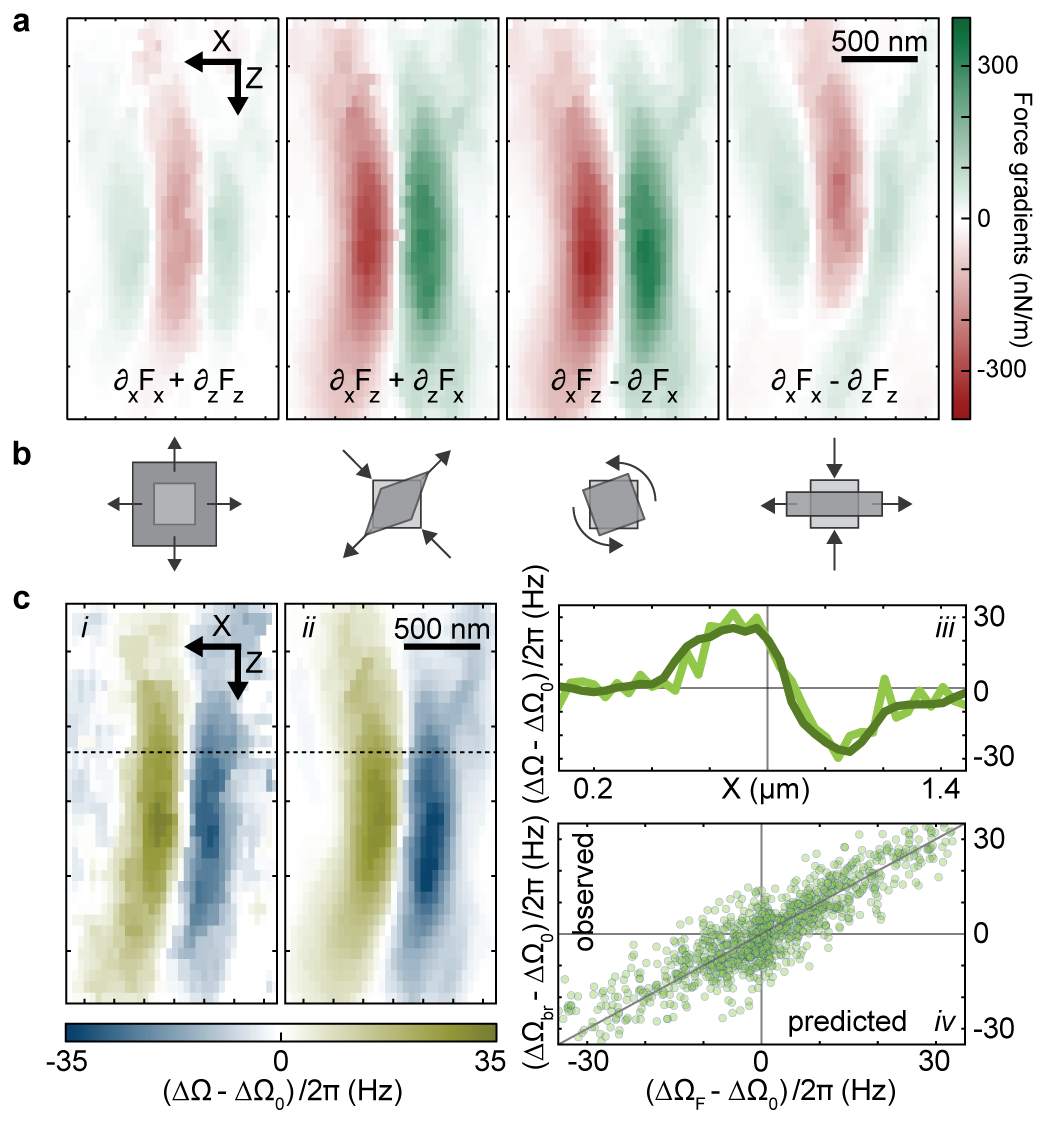}
\caption{
\textbf{Optomechanical backaction in two dimensions.} (a)  Projection of the force field derivatives $\partial_ i F_j $, $i,j\in\{x,z\}$ derived from Fig.\,2c on the real 2x2  Pauli matrices: $1, \sigma_x, i\sigma_y, \sigma_z$ which form a base of the force field gradients.
(e.g. $ i\sigma_y$=~\usebox{\smlmat} relates to $\partial_z F_x-\partial_xF_z$)
(b) Schematics of the force field topology associated to each base elements. The elementary deformation of a square unit cell permits visualizing  their divergence, shear and rotational properties. (c) Cartography of the relative frequency splitting $(\Delta\Omega-\Delta\Omega_0)/2\pi$:  (i) directly measured from Brownian motion measurements, (ii) predicted from (\ref{eq.splitting}) using the force field gradients of panel 3a after a proper rotation to account for eigenaxes orientations. Both measurements present a very good agreement as can be verified in the linear transverse cut (across the dotted line) shown in  (iii) (light green: direct, dark green: predicted) and in (iv) where all measurements are reported. The data points are well aligned on the first bisector (straight line), demonstrating that bidimensional optomechanical backaction governs the nanoresonator dynamics.}
\label{Fig3}
\end{center}
\end{figure}

This analysis therefore predicts that significant vectorial backaction can be expected in our system. Its direct consequence is a modification of the nanowire eigenfrequencies $\Omega_{1,2}\rightarrow\Omega_\pm\equiv \bar\Omega\pm\Delta\Omega/2$,  given by the square root of $\mathbf{\Omega^2}$ eigenvalues. In the limit of small frequency splitting $\Delta\Omega\equiv\Omega_+-\Omega_-$ with respect to the  mean frequency $\bar\Omega$ we have:
\begin{equation}
\Delta\Omega\equiv \frac{1}{2\bar\Omega}\sqrt{(\Omega_1^2-\Omega_2^2-g_{11}+g_{22})^2+4 g_{12}g_{21}}
\label{eq.splitting}
\end{equation}
whose expression captures all vectorial backaction effects. The consistency of our analysis can thus be tested by comparing the frequency splitting  $\Delta\Omega$ directly extracted from Brownian motion measurements (Fig. 3c.i), to the values inferred from (\ref{eq.splitting}) and the determination of the optical force field derivatives (Fig. 3a). Both approaches were found to be in perfect agreement (see Fig. 3c), underlying the necessity to have a full knowledge of force field topology in which the nanowire  evolves to properly understand the dynamical backaction  in 2D. \\
This backaction mechanism also depicts a novel class of strong coupling between both polarizations mediated by a bidimensional coupling mechanism, the eigenfrequencies $\Omega_{\pm}$ are shifted by more than their intrinsic linewidth $\Gamma$. This is an extension of the canonical scalar coupling between two harmonic oscillators, which efficiently describes the majority of strongly interacting systems and necessarily generates frequency repulsion  \cite{novotny2010strong}. This does not apply anymore in higher dimension, as can be seen in Fig. 3c where the frequency splitting $\Delta\Omega$ can be lower than its value in absence of light $\Delta\Omega_0$ .\\

\begin{figure*}[t!]
\begin{center}
\includegraphics[width= 0.9 \linewidth]{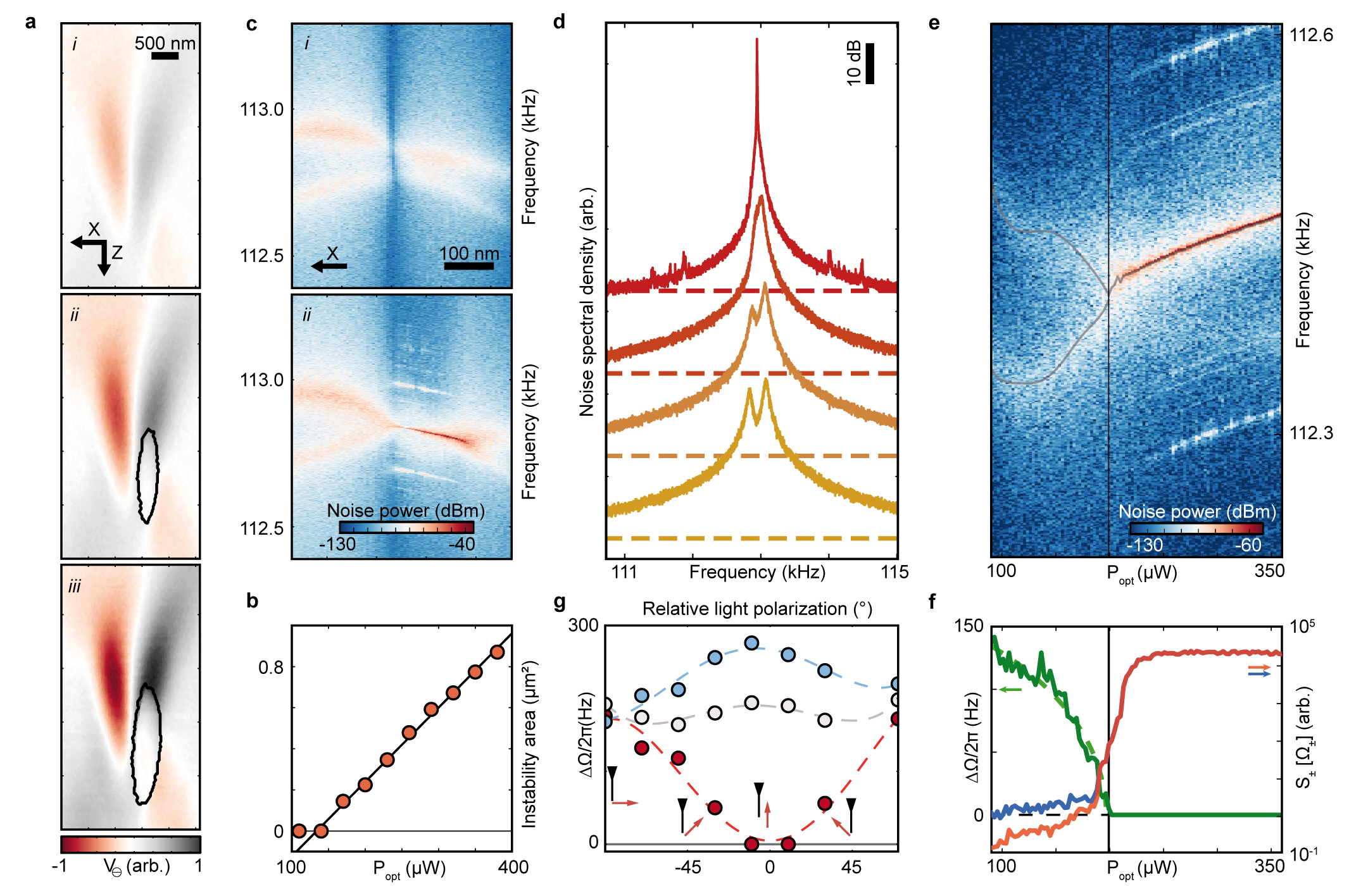}
\caption{
\textbf{: Dynamical bifurcation and topological instability in a non-conservative vectorial force field.} (a) Differential transmission maps measured for increasing optical powers (140, 260 and 380\,$\rm\mu W$ from {\it i} to {\it iii}, at 633\,nm). The contours of the topological instability region are stressed in black, its area increases with the incident optical power as shown in (b);  the solid line is a fit connected to the second order spatial derivatives of the force field (see SI). (c) Spatial dependence of  the displacement noise spectrum of the two fundamental eigenmodes,  measured while scanning the nanowire position across the optical axis for optical powers lower/larger (top/down) than the topological instability threshold. (d) Typical evolution of the displacement noise spectra when approaching and reaching the topological instability:  frequency pulling can be observed as well as the apparition of the coherent oscillation spike which is responsible for the blurring observed on DC transmission maps.
(e) Evolution of the displacement noise spectrum as a function of the incident optical power in the topologically unstable region. The corresponding frequency splittings  $\Delta\Omega/2\pi$ and peak noise powers $S_{\delta r_\beta}[\Omega_\pm]$ are reported in (f). The dashed line is a fit using eq.\,(\ref{eq.splitting}). (g) Frequency splitting as a function of the light polarization angle at constant incident power of $350\,\rm\mu W$, measured in three different locations (see SI).}
\label{Fig4}%
\end{center}
\end{figure*}

\textit{Bifurcation of the nanowire dynamics and topological instability.---}
At  larger optical powers, we could observe a direct and novel dynamical signature of the non-conservative topology of the 2D optomechanical interaction. Above a threshold of  $\sim 120\,\rm\mu W$, the nanoresonator starts to self-oscillate with extremely large amplitudes ($>500\,\rm nm$), sufficient to blur the DC transmission maps (Fig. 4a) in a specific area on the side of the optical axis, a region of large vorticity (Fig. 2d).
In the displacement noise spectra, one can observe an intense coherent peak growing on top of the thermal noise spectra, whose apparition is concomitant with a frequency merging of both eigenmodes (see Fig. 4c, 4d).
Indeed the shear terms product $g_{12}g_{21}$, which reflects the strength of cross-couplings between both polarizations and controls their frequency splitting $\Delta\Omega$, can accept negative values, except in the specific case of conservative force fields (then $g_{12}= g_{21}$). As a consequence, in regions of strong vorticity the frequency splitting defined in (\ref{eq.splitting}) can become purely imaginary at large optical powers, so that both eigenfrequencies are now complex conjugates $\Omega_\pm=\bar\Omega\pm i \rm Im(\Delta\Omega)/2$. The consequence is a frequency merging, as well as an asymmetric modification of the effective damping rates $\Gamma_\pm= \Gamma\pm \rm Im(\Delta\Omega)$.

The appearance of a coherent oscillation peak is observed when $\Gamma_-$ becomes negative and any force fluctuation can then initiate the dynamical topological instability. This bifurcation also corresponds to a transition from linearly to elliptically polarized eigenmodes,  rotating in opposite directions, so that the rotating force field transfers opposite work. The area of the topological instability -an ellipse just above threshold-  increases linearly with the optical power (see Fig.\,4.b), while the ellipse parameters are determined by the second order spatial derivatives of the force field (see SI). Its presence only on one side of the optical axis is due to the tilt of the eigendirections $\vv{e_{1,2}}$  with respect to the optical axis.\\
The bifurcation can also be investigated by placing the nanowire in the instable region and progressively increasing the optical power while recording the displacement noise spectra (see Fig.\,4e). Frequency merging is clearly visible as well as the onset of the topological instability. The measured frequency splitting $\Delta\Omega$ is reported in panel 4f as a function of optical power, showing a good qualitative agreement with a fit using expression (\ref{eq.splitting}) and  the experimentally measured force field gradients. In this measurement, the Q-factor is too large to directly discriminate frequency merging ($\Delta\Omega=0$)  from the onset of topological instability ($\Gamma_- = 0$).
However, increasing the mechanical damping rate $\Gamma$ through pressure increase at constant optical power beyond the bifurcation threshold leads to a reduction of the instability area, as expected from an increase of the instability threshold power (see SI).
Furthermore, the optical polarization offers another possibility to tune the optomechanical interaction:  the displacement readout signal-to-noise but also the light backaction are  thus maximized (see SI and Fig.\,4f) for vertical polarization, a situation in which the nanowire presents a larger optical polarisability \cite{Bohren1983,nichol2008displacement}. It is important to point out that the bifurcation described here does not involve non-linear dynamics and neither cavity-delayed optical forces such as in the well-known 1D optomechanical instability \cite{Kippenberg2008, Arcizet2006} or in recent studies on cavity-coupled multi-oscillator dynamics \cite{Heinrich2011,Zhang2012,Seok2013}.

Finally, the vectorial and scanning probe character of our nano-optomechanical approach represents a sensitive and robust strategy for further exploration of vectorial force fields at the nanoscale and is particularly suited for investigating the light-matter interaction in confined geometries, complementary to other characterization tools \cite{Bauer2013}.  Exploration of  optically resonant nano-optomechanical interactions related to internal Mie resonances of the nanowires \cite{Bohren1983} can be envisioned,  possibly leading to wavelength-dependent trapping or anti-trapping optical force fields but also negative scattering forces \cite{Chen2011}.\\
Our results demonstrate the importance of considering the full vectorial character of the optomechanical interaction to correctly describe the multimode thermodynamics of strongly coupled optomechanical systems \cite{Seok2013}. In the case of nano-optomechanics, noise contamination between eigenmodes can be anticipated as well as a violation of the fluctuation-dissipation theorem while noise thermometry can be significantly biased in spatially confined optomechanical systems \cite{Roichman2008,Sun2009}. A salient illustration can be seen in Fig.\,4e/4f, where the peak displacement noise presents a dramatic dependence on light intensity in regions of strong vorticity, increasing by more than 2 orders of magnitude before the bifurcation.
These considerations are  universal and crucial to the proper understanding of nanomechanics, when the resonator dimensions becomes smaller than the range of the force field in which they evolve.
As such, our system offers a unique platform to further investigate the fundamental connections between both acceptations of the term ``non-conservative" \cite{Krechetnikov2007}: the vorticity of a 2D force field in which the nanoresonator evolves and the dissipative character of the corresponding dynamics observed in 1D.\\

We thank A. Kuhn, B. Pigeau, G. Nogues, J.P. Poizat, N. Roch, S. Seidelin, J. Jarreau, L. Del Rey, D. Maillard, C. Hoarau and D. Lepoittevin  for fruitful interactions and technical developments. This work was supported by the Agence Nationale de la Recherche (RPDoc-2010, Blanc-2013-Focus, the European Research Council (ERC-StG-2012, HQ-NOM) and LANEF (CryOptics).

\bibliography{article_arnaud}

\begin{thebibliography}{57}
\expandafter\ifx\csname natexlab\endcsname\relax\def\natexlab#1{#1}\fi
\expandafter\ifx\csname bibnamefont\endcsname\relax
  \def\bibnamefont#1{#1}\fi
\expandafter\ifx\csname bibfnamefont\endcsname\relax
  \def\bibfnamefont#1{#1}\fi
\expandafter\ifx\csname citenamefont\endcsname\relax
  \def\citenamefont#1{#1}\fi
\expandafter\ifx\csname url\endcsname\relax
  \def\url#1{\texttt{#1}}\fi
\expandafter\ifx\csname urlprefix\endcsname\relax\def\urlprefix{URL }\fi
\providecommand{\bibinfo}[2]{#2}
\providecommand{\eprint}[2][]{\url{#2}}

\bibitem[{\citenamefont{Bohr}(1949)}]{bohr1949discussion}
\bibinfo{author}{\bibfnamefont{N.}~\bibnamefont{Bohr}},
  \emph{\bibinfo{title}{Discussion with Einstein on epistemological problems in
  atomic physics}} (\bibinfo{publisher}{University of Copenhagen},
  \bibinfo{year}{1949}).

\bibitem[{\citenamefont{Caves}(1982)}]{caves1982quantum}
\bibinfo{author}{\bibfnamefont{C.~M.} \bibnamefont{Caves}},
  \bibinfo{journal}{Physical Review D} \textbf{\bibinfo{volume}{26}},
  \bibinfo{pages}{1817} (\bibinfo{year}{1982}).

\bibitem[{\citenamefont{Braginsky et~al.}(1995)\citenamefont{Braginsky,
  Khalili, and Thorne}}]{braginsky1995quantum}
\bibinfo{author}{\bibfnamefont{V.~B.} \bibnamefont{Braginsky}},
  \bibinfo{author}{\bibfnamefont{F.~Y.} \bibnamefont{Khalili}},
  \bibnamefont{and} \bibinfo{author}{\bibfnamefont{K.~S.}
  \bibnamefont{Thorne}}, \emph{\bibinfo{title}{Quantum measurement}}
  (\bibinfo{publisher}{Cambridge University Press}, \bibinfo{year}{1995}).

\bibitem[{\citenamefont{Jaekel and Reynaud}(2007)}]{jaekel2007quantum}
\bibinfo{author}{\bibfnamefont{M.~T.} \bibnamefont{Jaekel}} \bibnamefont{and}
  \bibinfo{author}{\bibfnamefont{S.}~\bibnamefont{Reynaud}},
  \bibinfo{journal}{EPL (Europhysics Letters)} \textbf{\bibinfo{volume}{13}},
  \bibinfo{pages}{301} (\bibinfo{year}{2007}).

\bibitem[{\citenamefont{Dorsel et~al.}(1983)\citenamefont{Dorsel, McCullen,
  Meystre, Vignes, and Walther}}]{dorsel1983optical}
\bibinfo{author}{\bibfnamefont{A.}~\bibnamefont{Dorsel}},
  \bibinfo{author}{\bibfnamefont{J.}~\bibnamefont{McCullen}},
  \bibinfo{author}{\bibfnamefont{P.}~\bibnamefont{Meystre}},
  \bibinfo{author}{\bibfnamefont{E.}~\bibnamefont{Vignes}}, \bibnamefont{and}
  \bibinfo{author}{\bibfnamefont{H.}~\bibnamefont{Walther}},
  \bibinfo{journal}{Phys. Rev. Lett.} \textbf{\bibinfo{volume}{51}},
  \bibinfo{pages}{1550} (\bibinfo{year}{1983}).

\bibitem[{\citenamefont{Cohadon et~al.}(1999)\citenamefont{Cohadon, Heidmann,
  and Pinard}}]{Cohadon1999}
\bibinfo{author}{\bibfnamefont{P.~F.} \bibnamefont{Cohadon}},
  \bibinfo{author}{\bibfnamefont{A.}~\bibnamefont{Heidmann}}, \bibnamefont{and}
  \bibinfo{author}{\bibfnamefont{M.}~\bibnamefont{Pinard}},
  \bibinfo{journal}{Phys. Rev. Lett.} \textbf{\bibinfo{volume}{83}},
  \bibinfo{pages}{3174} (\bibinfo{year}{1999}).

\bibitem[{\citenamefont{Kippenberg and Vahala}(2008)}]{Kippenberg2008}
\bibinfo{author}{\bibfnamefont{T.~J.} \bibnamefont{Kippenberg}}
  \bibnamefont{and} \bibinfo{author}{\bibfnamefont{K.~J.}
  \bibnamefont{Vahala}}, \bibinfo{journal}{Science}
  \textbf{\bibinfo{volume}{321}}, \bibinfo{pages}{1172} (\bibinfo{year}{2008}).

\bibitem[{\citenamefont{Arcizet et~al.}(2006)\citenamefont{Arcizet, Cohadon,
  Briant, Pinard, and Heidmann}}]{Arcizet2006}
\bibinfo{author}{\bibfnamefont{O.}~\bibnamefont{Arcizet}},
  \bibinfo{author}{\bibfnamefont{P.-F.} \bibnamefont{Cohadon}},
  \bibinfo{author}{\bibfnamefont{T.}~\bibnamefont{Briant}},
  \bibinfo{author}{\bibfnamefont{M.}~\bibnamefont{Pinard}}, \bibnamefont{and}
  \bibinfo{author}{\bibfnamefont{A.}~\bibnamefont{Heidmann}},
  \bibinfo{journal}{Nature} \textbf{\bibinfo{volume}{444}}, \bibinfo{pages}{71}
  (\bibinfo{year}{2006}).

\bibitem[{\citenamefont{Gigan et~al.}(2006)\citenamefont{Gigan, B\"ohm,
  Paternostro, Blaser, Langer, Hertzberg, Schwab, B\"auerle, Aspelmeyer, and
  Zeilinger}}]{Gigan2006}
\bibinfo{author}{\bibfnamefont{S.}~\bibnamefont{Gigan}},
  \bibinfo{author}{\bibfnamefont{H.~R.} \bibnamefont{B\"ohm}},
  \bibinfo{author}{\bibfnamefont{M.}~\bibnamefont{Paternostro}},
  \bibinfo{author}{\bibfnamefont{F.}~\bibnamefont{Blaser}},
  \bibinfo{author}{\bibfnamefont{G.}~\bibnamefont{Langer}},
  \bibinfo{author}{\bibfnamefont{J.~B.} \bibnamefont{Hertzberg}},
  \bibinfo{author}{\bibfnamefont{K.~C.} \bibnamefont{Schwab}},
  \bibinfo{author}{\bibfnamefont{D.}~\bibnamefont{B\"auerle}},
  \bibinfo{author}{\bibfnamefont{M.}~\bibnamefont{Aspelmeyer}},
  \bibnamefont{and}
  \bibinfo{author}{\bibfnamefont{A.}~\bibnamefont{Zeilinger}},
  \bibinfo{journal}{Nature} \textbf{\bibinfo{volume}{444}}, \bibinfo{pages}{67}
  (\bibinfo{year}{2006}).

\bibitem[{\citenamefont{Thompson et~al.}(2008)\citenamefont{Thompson, Zwickl,
  Jayich, Marquardt, Girvin, and Harris}}]{Thomson2008}
\bibinfo{author}{\bibfnamefont{J.~D.} \bibnamefont{Thompson}},
  \bibinfo{author}{\bibfnamefont{B.~M.} \bibnamefont{Zwickl}},
  \bibinfo{author}{\bibfnamefont{A.~M.} \bibnamefont{Jayich}},
  \bibinfo{author}{\bibfnamefont{F.}~\bibnamefont{Marquardt}},
  \bibinfo{author}{\bibfnamefont{S.~M.} \bibnamefont{Girvin}},
  \bibnamefont{and} \bibinfo{author}{\bibfnamefont{J.~G.~E.}
  \bibnamefont{Harris}}, \bibinfo{journal}{Nature}
  \textbf{\bibinfo{volume}{452}} (\bibinfo{year}{2008}).

\bibitem[{\citenamefont{Teufel et~al.}(2011)\citenamefont{Teufel, Donner, Li,
  Harlow, Allman, Cicak, Sirois, Whittaker, Lehnert, and
  Simmonds}}]{Teufel2011}
\bibinfo{author}{\bibfnamefont{J.~D.} \bibnamefont{Teufel}},
  \bibinfo{author}{\bibfnamefont{T.}~\bibnamefont{Donner}},
  \bibinfo{author}{\bibfnamefont{D.}~\bibnamefont{Li}},
  \bibinfo{author}{\bibfnamefont{J.~W.} \bibnamefont{Harlow}},
  \bibinfo{author}{\bibfnamefont{M.~S.} \bibnamefont{Allman}},
  \bibinfo{author}{\bibfnamefont{K.}~\bibnamefont{Cicak}},
  \bibinfo{author}{\bibfnamefont{A.~J.} \bibnamefont{Sirois}},
  \bibinfo{author}{\bibfnamefont{J.~D.} \bibnamefont{Whittaker}},
  \bibinfo{author}{\bibfnamefont{K.~W.} \bibnamefont{Lehnert}},
  \bibnamefont{and} \bibinfo{author}{\bibfnamefont{R.~W.}
  \bibnamefont{Simmonds}}, \bibinfo{journal}{Nature}
  \textbf{\bibinfo{volume}{475}}, \bibinfo{pages}{359} (\bibinfo{year}{2011}).

\bibitem[{\citenamefont{Chan et~al.}(2011)\citenamefont{Chan, Alegre,
  Safavi-Naeini, Hill, Krause, Gr{\"o}blacher, Aspelmeyer, and
  Painter}}]{chan2011laser}
\bibinfo{author}{\bibfnamefont{J.}~\bibnamefont{Chan}},
  \bibinfo{author}{\bibfnamefont{T.}~\bibnamefont{Alegre}},
  \bibinfo{author}{\bibfnamefont{A.}~\bibnamefont{Safavi-Naeini}},
  \bibinfo{author}{\bibfnamefont{J.}~\bibnamefont{Hill}},
  \bibinfo{author}{\bibfnamefont{A.}~\bibnamefont{Krause}},
  \bibinfo{author}{\bibfnamefont{S.}~\bibnamefont{Gr{\"o}blacher}},
  \bibinfo{author}{\bibfnamefont{M.}~\bibnamefont{Aspelmeyer}},
  \bibnamefont{and} \bibinfo{author}{\bibfnamefont{O.}~\bibnamefont{Painter}},
  \bibinfo{journal}{Nature} \textbf{\bibinfo{volume}{478}}, \bibinfo{pages}{89}
  (\bibinfo{year}{2011}).

\bibitem[{\citenamefont{Verhagen et~al.}(2012)\citenamefont{Verhagen,
  Deléglise, Weis, Schliesser, and Kippenberg}}]{Verhagen2012}
\bibinfo{author}{\bibfnamefont{E.}~\bibnamefont{Verhagen}},
  \bibinfo{author}{\bibfnamefont{S.}~\bibnamefont{Deléglise}},
  \bibinfo{author}{\bibfnamefont{S.}~\bibnamefont{Weis}},
  \bibinfo{author}{\bibfnamefont{A.}~\bibnamefont{Schliesser}},
  \bibnamefont{and}
  \bibinfo{author}{\bibfnamefont{T.}~\bibnamefont{Kippenberg}},
  \bibinfo{journal}{Nature} \textbf{\bibinfo{volume}{4824}},
  \bibinfo{pages}{63} (\bibinfo{year}{2012}).

\bibitem[{\citenamefont{Verlot et~al.}(2009)\citenamefont{Verlot, Tavernarakis,
  Briant, Cohadon, and Heidmann}}]{verlot2009}
\bibinfo{author}{\bibfnamefont{P.}~\bibnamefont{Verlot}},
  \bibinfo{author}{\bibfnamefont{A.}~\bibnamefont{Tavernarakis}},
  \bibinfo{author}{\bibfnamefont{T.}~\bibnamefont{Briant}},
  \bibinfo{author}{\bibfnamefont{P.}~\bibnamefont{Cohadon}}, \bibnamefont{and}
  \bibinfo{author}{\bibfnamefont{A.}~\bibnamefont{Heidmann}},
  \bibinfo{journal}{Phys. Rev. Lett.} \textbf{\bibinfo{volume}{102}},
  \bibinfo{pages}{103601} (\bibinfo{year}{2009}).

\bibitem[{\citenamefont{Marino et~al.}(2010)\citenamefont{Marino, Cataliotti,
  Farsi, de~Cumis, and Marin}}]{marino2010classical}
\bibinfo{author}{\bibfnamefont{F.}~\bibnamefont{Marino}},
  \bibinfo{author}{\bibfnamefont{F.~S.} \bibnamefont{Cataliotti}},
  \bibinfo{author}{\bibfnamefont{A.}~\bibnamefont{Farsi}},
  \bibinfo{author}{\bibfnamefont{M.~S.} \bibnamefont{de~Cumis}},
  \bibnamefont{and} \bibinfo{author}{\bibfnamefont{F.}~\bibnamefont{Marin}},
  \bibinfo{journal}{Phys. Rev. Lett.} \textbf{\bibinfo{volume}{104}},
  \bibinfo{pages}{73601} (\bibinfo{year}{2010}).

\bibitem[{\citenamefont{Purdy et~al.}(2013)\citenamefont{Purdy, Peterson, and
  Regal}}]{purdy2013observation}
\bibinfo{author}{\bibfnamefont{T.}~\bibnamefont{Purdy}},
  \bibinfo{author}{\bibfnamefont{R.}~\bibnamefont{Peterson}}, \bibnamefont{and}
  \bibinfo{author}{\bibfnamefont{C.}~\bibnamefont{Regal}},
  \bibinfo{journal}{Science} \textbf{\bibinfo{volume}{339}},
  \bibinfo{pages}{801} (\bibinfo{year}{2013}).

\bibitem[{\citenamefont{Verlot et~al.}(2010)\citenamefont{Verlot, Tavernarakis,
  Briant, Cohadon, and Heidmann}}]{verlot2010backaction}
\bibinfo{author}{\bibfnamefont{P.}~\bibnamefont{Verlot}},
  \bibinfo{author}{\bibfnamefont{A.}~\bibnamefont{Tavernarakis}},
  \bibinfo{author}{\bibfnamefont{T.}~\bibnamefont{Briant}},
  \bibinfo{author}{\bibfnamefont{P.}~\bibnamefont{Cohadon}}, \bibnamefont{and}
  \bibinfo{author}{\bibfnamefont{A.}~\bibnamefont{Heidmann}},
  \bibinfo{journal}{Phys. Rev. Lett.} \textbf{\bibinfo{volume}{104}},
  \bibinfo{pages}{133602} (\bibinfo{year}{2010}).

\bibitem[{\citenamefont{Weis et~al.}(2010)\citenamefont{Weis, Rivi\`ere,
  Del\'eglise, Gavartin, Arcizet, Schliesser, and Kippenberg}}]{Weis2010}
\bibinfo{author}{\bibfnamefont{S.}~\bibnamefont{Weis}},
  \bibinfo{author}{\bibfnamefont{R.}~\bibnamefont{Rivi\`ere}},
  \bibinfo{author}{\bibfnamefont{S.}~\bibnamefont{Del\'eglise}},
  \bibinfo{author}{\bibfnamefont{E.}~\bibnamefont{Gavartin}},
  \bibinfo{author}{\bibfnamefont{O.}~\bibnamefont{Arcizet}},
  \bibinfo{author}{\bibfnamefont{A.}~\bibnamefont{Schliesser}},
  \bibnamefont{and} \bibinfo{author}{\bibfnamefont{T.~J.}
  \bibnamefont{Kippenberg}}, \bibinfo{journal}{Science}
  \textbf{\bibinfo{volume}{330}}, \bibinfo{pages}{1520} (\bibinfo{year}{2010}).

\bibitem[{\citenamefont{Safavi-Naeini et~al.}(2011)\citenamefont{Safavi-Naeini,
  Alegre, Chan, Eichenfield, Winger, Lin, Hill, Chang, and
  Painter}}]{safavi2011electromagnetically}
\bibinfo{author}{\bibfnamefont{A.~H.} \bibnamefont{Safavi-Naeini}},
  \bibinfo{author}{\bibfnamefont{T.~M.} \bibnamefont{Alegre}},
  \bibinfo{author}{\bibfnamefont{J.}~\bibnamefont{Chan}},
  \bibinfo{author}{\bibfnamefont{M.}~\bibnamefont{Eichenfield}},
  \bibinfo{author}{\bibfnamefont{M.}~\bibnamefont{Winger}},
  \bibinfo{author}{\bibfnamefont{Q.}~\bibnamefont{Lin}},
  \bibinfo{author}{\bibfnamefont{J.~T.} \bibnamefont{Hill}},
  \bibinfo{author}{\bibfnamefont{D.}~\bibnamefont{Chang}}, \bibnamefont{and}
  \bibinfo{author}{\bibfnamefont{O.}~\bibnamefont{Painter}},
  \bibinfo{journal}{Nature} \textbf{\bibinfo{volume}{472}}, \bibinfo{pages}{69}
  (\bibinfo{year}{2011}).

\bibitem[{\citenamefont{Pinard et~al.}(1999)\citenamefont{Pinard, Hadjar, and
  Heidmann}}]{Pinard1999}
\bibinfo{author}{\bibfnamefont{M.}~\bibnamefont{Pinard}},
  \bibinfo{author}{\bibfnamefont{Y.}~\bibnamefont{Hadjar}}, \bibnamefont{and}
  \bibinfo{author}{\bibfnamefont{A.}~\bibnamefont{Heidmann}},
  \bibinfo{journal}{Eur. Phys. J. D} \textbf{\bibinfo{volume}{7}},
  \bibinfo{pages}{107} (\bibinfo{year}{1999}).

\bibitem[{\citenamefont{Ashkin et~al.}(1986)\citenamefont{Ashkin, Dziedzic,
  Bjorkholm, and Chu}}]{Ashkin1986}
\bibinfo{author}{\bibfnamefont{A.}~\bibnamefont{Ashkin}},
  \bibinfo{author}{\bibfnamefont{J.~M.} \bibnamefont{Dziedzic}},
  \bibinfo{author}{\bibfnamefont{J.~E.} \bibnamefont{Bjorkholm}},
  \bibnamefont{and} \bibinfo{author}{\bibfnamefont{S.}~\bibnamefont{Chu}},
  \bibinfo{journal}{Opt. Lett.} \textbf{\bibinfo{volume}{11}},
  \bibinfo{pages}{288} (\bibinfo{year}{1986}).

\bibitem[{\citenamefont{Cohen-Tannoudji
  et~al.}(2004)\citenamefont{Cohen-Tannoudji, Dupont-Roc, and
  Grynberg}}]{Cohen-Tannoudji2004}
\bibinfo{author}{\bibfnamefont{C.}~\bibnamefont{Cohen-Tannoudji}},
  \bibinfo{author}{\bibfnamefont{J.}~\bibnamefont{Dupont-Roc}},
  \bibnamefont{and} \bibinfo{author}{\bibfnamefont{G.}~\bibnamefont{Grynberg}},
  \emph{\bibinfo{title}{Photons and Atoms: Introduction to Quantum
  Electrodynamics}} (\bibinfo{publisher}{Wiley}, \bibinfo{year}{2004}).

\bibitem[{\citenamefont{Novotny and Hecht}(2006)}]{Novotny2006}
\bibinfo{author}{\bibfnamefont{L.}~\bibnamefont{Novotny}} \bibnamefont{and}
  \bibinfo{author}{\bibfnamefont{B.}~\bibnamefont{Hecht}},
  \emph{\bibinfo{title}{Principles of Nano-Optics}}
  (\bibinfo{publisher}{Cambridge University Press}, \bibinfo{year}{2006}).

\bibitem[{\citenamefont{Roichman et~al.}(2008)\citenamefont{Roichman, Sun,
  Stolarski, and Grier}}]{Roichman2008}
\bibinfo{author}{\bibfnamefont{Y.}~\bibnamefont{Roichman}},
  \bibinfo{author}{\bibfnamefont{B.}~\bibnamefont{Sun}},
  \bibinfo{author}{\bibfnamefont{A.}~\bibnamefont{Stolarski}},
  \bibnamefont{and} \bibinfo{author}{\bibfnamefont{D.~G.} \bibnamefont{Grier}},
  \bibinfo{journal}{Phys. Rev. Lett.} \textbf{\bibinfo{volume}{101}},
  \bibinfo{pages}{128301} (\bibinfo{year}{2008}).

\bibitem[{\citenamefont{Sun et~al.}(2009)\citenamefont{Sun, Lin, Darby,
  Grosberg, and Grier}}]{Sun2009}
\bibinfo{author}{\bibfnamefont{B.}~\bibnamefont{Sun}},
  \bibinfo{author}{\bibfnamefont{J.}~\bibnamefont{Lin}},
  \bibinfo{author}{\bibfnamefont{E.}~\bibnamefont{Darby}},
  \bibinfo{author}{\bibfnamefont{A.~Y.} \bibnamefont{Grosberg}},
  \bibnamefont{and} \bibinfo{author}{\bibfnamefont{D.~G.} \bibnamefont{Grier}},
  \bibinfo{journal}{Phys. Rev. E} \textbf{\bibinfo{volume}{80}},
  \bibinfo{pages}{010401} (\bibinfo{year}{2009}).

\bibitem[{\citenamefont{Krechetnikov and Marsden}(2007)}]{Krechetnikov2007}
\bibinfo{author}{\bibfnamefont{R.}~\bibnamefont{Krechetnikov}}
  \bibnamefont{and} \bibinfo{author}{\bibfnamefont{J.~E.}
  \bibnamefont{Marsden}}, \bibinfo{journal}{Rev. Mod. Phys.}
  \textbf{\bibinfo{volume}{79}}, \bibinfo{pages}{519} (\bibinfo{year}{2007}).

\bibitem[{\citenamefont{Li et~al.}(2008)\citenamefont{Li, Pernice, Xiong,
  Baehr-Jones, Hochberg, and Tang}}]{Li2008}
\bibinfo{author}{\bibfnamefont{M.}~\bibnamefont{Li}},
  \bibinfo{author}{\bibfnamefont{W.~H.~P.} \bibnamefont{Pernice}},
  \bibinfo{author}{\bibfnamefont{C.}~\bibnamefont{Xiong}},
  \bibinfo{author}{\bibfnamefont{T.}~\bibnamefont{Baehr-Jones}},
  \bibinfo{author}{\bibfnamefont{M.}~\bibnamefont{Hochberg}}, \bibnamefont{and}
  \bibinfo{author}{\bibfnamefont{H.~X.} \bibnamefont{Tang}},
  \bibinfo{journal}{Nature} \textbf{\bibinfo{volume}{456}},
  \bibinfo{pages}{480} (\bibinfo{year}{2008}).

\bibitem[{\citenamefont{Anetsberger et~al.}(2009)\citenamefont{Anetsberger,
  Arcizet, Unterreithmeier, Rivi\`ere, Schliesser, Weig, Kotthaus, and
  Kippenberg}}]{Anetsberger2009}
\bibinfo{author}{\bibfnamefont{G.}~\bibnamefont{Anetsberger}},
  \bibinfo{author}{\bibfnamefont{O.}~\bibnamefont{Arcizet}},
  \bibinfo{author}{\bibfnamefont{Q.~P.} \bibnamefont{Unterreithmeier}},
  \bibinfo{author}{\bibfnamefont{R.}~\bibnamefont{Rivi\`ere}},
  \bibinfo{author}{\bibfnamefont{A.}~\bibnamefont{Schliesser}},
  \bibinfo{author}{\bibfnamefont{E.~M.} \bibnamefont{Weig}},
  \bibinfo{author}{\bibfnamefont{J.~P.} \bibnamefont{Kotthaus}},
  \bibnamefont{and} \bibinfo{author}{\bibfnamefont{T.~J.}
  \bibnamefont{Kippenberg}}, \bibinfo{journal}{Nature Physics}
  \textbf{\bibinfo{volume}{5}}, \bibinfo{pages}{909} (\bibinfo{year}{2009}).

\bibitem[{\citenamefont{Favero and Karrai}(2009)}]{Favero2009}
\bibinfo{author}{\bibfnamefont{I.}~\bibnamefont{Favero}} \bibnamefont{and}
  \bibinfo{author}{\bibfnamefont{K.}~\bibnamefont{Karrai}},
  \bibinfo{journal}{Nature Photon.} \textbf{\bibinfo{volume}{3}},
  \bibinfo{pages}{201} (\bibinfo{year}{2009}).

\bibitem[{\citenamefont{Gavartin et~al.}(2011)\citenamefont{Gavartin, Braive,
  Sagnes, Arcizet, Beveratos, Kippenberg, and Robert-Philip}}]{Gavartin2011}
\bibinfo{author}{\bibfnamefont{E.}~\bibnamefont{Gavartin}},
  \bibinfo{author}{\bibfnamefont{R.}~\bibnamefont{Braive}},
  \bibinfo{author}{\bibfnamefont{I.}~\bibnamefont{Sagnes}},
  \bibinfo{author}{\bibfnamefont{O.}~\bibnamefont{Arcizet}},
  \bibinfo{author}{\bibfnamefont{A.}~\bibnamefont{Beveratos}},
  \bibinfo{author}{\bibfnamefont{T.~J.} \bibnamefont{Kippenberg}},
  \bibnamefont{and}
  \bibinfo{author}{\bibfnamefont{I.}~\bibnamefont{Robert-Philip}},
  \bibinfo{journal}{Phys. Rev. Lett.} \textbf{\bibinfo{volume}{106}},
  \bibinfo{pages}{203902} (\bibinfo{year}{2011}).

\bibitem[{\citenamefont{Ramos et~al.}(2012)\citenamefont{Ramos, Gil-Santos,
  Pini, Llorens, Fernández-Regúlez, San~Paulo, Calleja, and
  Tamayo}}]{Ramos2012}
\bibinfo{author}{\bibfnamefont{D.}~\bibnamefont{Ramos}},
  \bibinfo{author}{\bibfnamefont{E.}~\bibnamefont{Gil-Santos}},
  \bibinfo{author}{\bibfnamefont{V.}~\bibnamefont{Pini}},
  \bibinfo{author}{\bibfnamefont{J.~M.} \bibnamefont{Llorens}},
  \bibinfo{author}{\bibfnamefont{M.}~\bibnamefont{Fernández-Regúlez}},
  \bibinfo{author}{\bibfnamefont{Ã.}~\bibnamefont{San~Paulo}},
  \bibinfo{author}{\bibfnamefont{M.}~\bibnamefont{Calleja}}, \bibnamefont{and}
  \bibinfo{author}{\bibfnamefont{J.}~\bibnamefont{Tamayo}},
  \bibinfo{journal}{Nano Letters} \textbf{\bibinfo{volume}{12}},
  \bibinfo{pages}{932} (\bibinfo{year}{2012}).

\bibitem[{\citenamefont{Perisanu et~al.}(2007)\citenamefont{Perisanu, Vincent,
  Ayari, Choueib, Purcell, Bechelany, and Cornu}}]{Perisanu2007}
\bibinfo{author}{\bibfnamefont{S.}~\bibnamefont{Perisanu}},
  \bibinfo{author}{\bibfnamefont{P.}~\bibnamefont{Vincent}},
  \bibinfo{author}{\bibfnamefont{A.}~\bibnamefont{Ayari}},
  \bibinfo{author}{\bibfnamefont{M.}~\bibnamefont{Choueib}},
  \bibinfo{author}{\bibfnamefont{S.~T.} \bibnamefont{Purcell}},
  \bibinfo{author}{\bibfnamefont{M.}~\bibnamefont{Bechelany}},
  \bibnamefont{and} \bibinfo{author}{\bibfnamefont{D.}~\bibnamefont{Cornu}},
  \bibinfo{journal}{Applied Physics Letters} \textbf{\bibinfo{volume}{90}},
  \bibinfo{eid}{043113} (\bibinfo{year}{2007}).

\bibitem[{\citenamefont{Ren et~al.}(1996)\citenamefont{Ren, Gr{\'e}han, and
  Gouesbet}}]{ren1996prediction}
\bibinfo{author}{\bibfnamefont{K.}~\bibnamefont{Ren}},
  \bibinfo{author}{\bibfnamefont{G.}~\bibnamefont{Gr{\'e}han}},
  \bibnamefont{and} \bibinfo{author}{\bibfnamefont{G.}~\bibnamefont{Gouesbet}},
  \bibinfo{journal}{Applied optics} \textbf{\bibinfo{volume}{35}},
  \bibinfo{pages}{2702} (\bibinfo{year}{1996}).

\bibitem[{\citenamefont{Dogariu et~al.}(2012)\citenamefont{Dogariu, Sukhov, and
  S{\'a}enz}}]{dogariu2012optically}
\bibinfo{author}{\bibfnamefont{A.}~\bibnamefont{Dogariu}},
  \bibinfo{author}{\bibfnamefont{S.}~\bibnamefont{Sukhov}}, \bibnamefont{and}
  \bibinfo{author}{\bibfnamefont{J.}~\bibnamefont{S{\'a}enz}},
  \bibinfo{journal}{Nature Photonics} \textbf{\bibinfo{volume}{7}},
  \bibinfo{pages}{24} (\bibinfo{year}{2012}).

\bibitem[{\citenamefont{Wilson-Rae et~al.}(2004)\citenamefont{Wilson-Rae,
  Zoller, and Imamoglu}}]{Wilson-Rae2004}
\bibinfo{author}{\bibfnamefont{I.}~\bibnamefont{Wilson-Rae}},
  \bibinfo{author}{\bibfnamefont{P.}~\bibnamefont{Zoller}}, \bibnamefont{and}
  \bibinfo{author}{\bibfnamefont{A.}~\bibnamefont{Imamoglu}},
  \bibinfo{journal}{Physical Review Letters} \textbf{\bibinfo{volume}{92}},
  \bibinfo{pages}{075507} (\bibinfo{year}{2004}).

\bibitem[{\citenamefont{Rabl et~al.}(2010)\citenamefont{Rabl, Kolkowitz,
  Koppens, Harris, Zoller, and Lukin}}]{rabl2010quantum}
\bibinfo{author}{\bibfnamefont{P.}~\bibnamefont{Rabl}},
  \bibinfo{author}{\bibfnamefont{S.}~\bibnamefont{Kolkowitz}},
  \bibinfo{author}{\bibfnamefont{F.}~\bibnamefont{Koppens}},
  \bibinfo{author}{\bibfnamefont{J.}~\bibnamefont{Harris}},
  \bibinfo{author}{\bibfnamefont{P.}~\bibnamefont{Zoller}}, \bibnamefont{and}
  \bibinfo{author}{\bibfnamefont{M.}~\bibnamefont{Lukin}},
  \bibinfo{journal}{Nature Physics} \textbf{\bibinfo{volume}{6}},
  \bibinfo{pages}{602} (\bibinfo{year}{2010}).

\bibitem[{\citenamefont{Arcizet et~al.}(2011)\citenamefont{Arcizet, Jacques,
  Siria, Poncharal, Vincent, and Seidelin}}]{Arcizet2011}
\bibinfo{author}{\bibfnamefont{O.}~\bibnamefont{Arcizet}},
  \bibinfo{author}{\bibfnamefont{V.}~\bibnamefont{Jacques}},
  \bibinfo{author}{\bibfnamefont{A.}~\bibnamefont{Siria}},
  \bibinfo{author}{\bibfnamefont{P.}~\bibnamefont{Poncharal}},
  \bibinfo{author}{\bibfnamefont{P.}~\bibnamefont{Vincent}}, \bibnamefont{and}
  \bibinfo{author}{\bibfnamefont{S.}~\bibnamefont{Seidelin}},
  \bibinfo{journal}{Nature Phys.} \textbf{\bibinfo{volume}{7}},
  \bibinfo{pages}{879} (\bibinfo{year}{2011}).

\bibitem[{\citenamefont{Yeo et~al.}(2013)\citenamefont{Yeo, de~Assis, Gloppe,
  Dupont-Ferrier, Verlot, Malik, Dupuy, Claudon, G\'{e}rard, Auff\`{e}ves
  et~al.}}]{Yeo2013}
\bibinfo{author}{\bibfnamefont{I.}~\bibnamefont{Yeo}},
  \bibinfo{author}{\bibfnamefont{P.}~\bibnamefont{de~Assis}},
  \bibinfo{author}{\bibfnamefont{A.}~\bibnamefont{Gloppe}},
  \bibinfo{author}{\bibfnamefont{E.}~\bibnamefont{Dupont-Ferrier}},
  \bibinfo{author}{\bibfnamefont{P.}~\bibnamefont{Verlot}},
  \bibinfo{author}{\bibfnamefont{N.}~\bibnamefont{Malik}},
  \bibinfo{author}{\bibfnamefont{E.}~\bibnamefont{Dupuy}},
  \bibinfo{author}{\bibfnamefont{J.}~\bibnamefont{Claudon}},
  \bibinfo{author}{\bibfnamefont{J.}~\bibnamefont{G\'{e}rard}},
  \bibinfo{author}{\bibfnamefont{A.}~\bibnamefont{Auff\`{e}ves}},
  \bibnamefont{et~al.}, \bibinfo{journal}{Nature Nano., arXiv:1306.4209}
  (\bibinfo{year}{2013}).

\bibitem[{\citenamefont{Gavartin et~al.}(2012)\citenamefont{Gavartin, Verlot,
  and Kippenberg}}]{Gavartin2012}
\bibinfo{author}{\bibfnamefont{E.}~\bibnamefont{Gavartin}},
  \bibinfo{author}{\bibfnamefont{P.}~\bibnamefont{Verlot}}, \bibnamefont{and}
  \bibinfo{author}{\bibfnamefont{T.}~\bibnamefont{Kippenberg}},
  \bibinfo{journal}{Nature Nanotech.} \textbf{\bibinfo{volume}{7}},
  \bibinfo{pages}{509} (\bibinfo{year}{2012}).

\bibitem[{\citenamefont{Chaste et~al.}(2012)\citenamefont{Chaste, Eichler,
  Moser, Ceballos, Rurali, and Bachtold}}]{Chaste2012}
\bibinfo{author}{\bibfnamefont{J.}~\bibnamefont{Chaste}},
  \bibinfo{author}{\bibfnamefont{A.}~\bibnamefont{Eichler}},
  \bibinfo{author}{\bibfnamefont{J.}~\bibnamefont{Moser}},
  \bibinfo{author}{\bibfnamefont{G.}~\bibnamefont{Ceballos}},
  \bibinfo{author}{\bibfnamefont{R.}~\bibnamefont{Rurali}}, \bibnamefont{and}
  \bibinfo{author}{\bibfnamefont{A.}~\bibnamefont{Bachtold}},
  \bibinfo{journal}{Nature Nanotech.} \textbf{\bibinfo{volume}{7}},
  \bibinfo{pages}{300} (\bibinfo{year}{2012}).

\bibitem[{\citenamefont{Hanay et~al.}(2012)\citenamefont{Hanay, Kelber, Naik,
  Chi, Hentz, Bullard, Colinet, Duraffourg, and Roukes}}]{hanay2012single}
\bibinfo{author}{\bibfnamefont{M.}~\bibnamefont{Hanay}},
  \bibinfo{author}{\bibfnamefont{S.}~\bibnamefont{Kelber}},
  \bibinfo{author}{\bibfnamefont{A.}~\bibnamefont{Naik}},
  \bibinfo{author}{\bibfnamefont{D.}~\bibnamefont{Chi}},
  \bibinfo{author}{\bibfnamefont{S.}~\bibnamefont{Hentz}},
  \bibinfo{author}{\bibfnamefont{E.}~\bibnamefont{Bullard}},
  \bibinfo{author}{\bibfnamefont{E.}~\bibnamefont{Colinet}},
  \bibinfo{author}{\bibfnamefont{L.}~\bibnamefont{Duraffourg}},
  \bibnamefont{and} \bibinfo{author}{\bibfnamefont{M.}~\bibnamefont{Roukes}},
  \bibinfo{journal}{Nature Nanotechnology} \textbf{\bibinfo{volume}{7}},
  \bibinfo{pages}{602} (\bibinfo{year}{2012}).

\bibitem[{\citenamefont{Nichol et~al.}(2012)\citenamefont{Nichol, Hemesath,
  Lauhon, and Budakian}}]{nichol2012viewpoint}
\bibinfo{author}{\bibfnamefont{J.}~\bibnamefont{Nichol}},
  \bibinfo{author}{\bibfnamefont{E.}~\bibnamefont{Hemesath}},
  \bibinfo{author}{\bibfnamefont{L.}~\bibnamefont{Lauhon}}, \bibnamefont{and}
  \bibinfo{author}{\bibfnamefont{R.}~\bibnamefont{Budakian}},
  \bibinfo{journal}{Phys. Rev. B} \textbf{\bibinfo{volume}{85}},
  \bibinfo{pages}{054414} (\bibinfo{year}{2012}).

\bibitem[{\citenamefont{Li et~al.}(2011)\citenamefont{Li, Kheifets, and
  Raizen}}]{Raizen2011}
\bibinfo{author}{\bibfnamefont{T.}~\bibnamefont{Li}},
  \bibinfo{author}{\bibfnamefont{S.}~\bibnamefont{Kheifets}}, \bibnamefont{and}
  \bibinfo{author}{\bibfnamefont{M.}~\bibnamefont{Raizen}},
  \bibinfo{journal}{Nature Phys.} \textbf{\bibinfo{volume}{7}},
  \bibinfo{pages}{527} (\bibinfo{year}{2011}).

\bibitem[{\citenamefont{Gieseler et~al.}(2012)\citenamefont{Gieseler, Deutsch,
  Quidant, and Novotny}}]{Gieseler2012}
\bibinfo{author}{\bibfnamefont{J.}~\bibnamefont{Gieseler}},
  \bibinfo{author}{\bibfnamefont{B.}~\bibnamefont{Deutsch}},
  \bibinfo{author}{\bibfnamefont{R.}~\bibnamefont{Quidant}}, \bibnamefont{and}
  \bibinfo{author}{\bibfnamefont{L.}~\bibnamefont{Novotny}},
  \bibinfo{journal}{Phys. Rev. Lett.} \textbf{\bibinfo{volume}{109}},
  \bibinfo{pages}{103603} (\bibinfo{year}{2012}).

\bibitem[{\citenamefont{Treps et~al.}(2003)\citenamefont{Treps, Grosse, Bowen,
  Fabre, Bachor, and Lam}}]{treps2003quantum}
\bibinfo{author}{\bibfnamefont{N.}~\bibnamefont{Treps}},
  \bibinfo{author}{\bibfnamefont{N.}~\bibnamefont{Grosse}},
  \bibinfo{author}{\bibfnamefont{W.~P.} \bibnamefont{Bowen}},
  \bibinfo{author}{\bibfnamefont{C.}~\bibnamefont{Fabre}},
  \bibinfo{author}{\bibfnamefont{H.-A.} \bibnamefont{Bachor}},
  \bibnamefont{and} \bibinfo{author}{\bibfnamefont{P.~K.} \bibnamefont{Lam}},
  \bibinfo{journal}{Science} \textbf{\bibinfo{volume}{301}},
  \bibinfo{pages}{940} (\bibinfo{year}{2003}).

\bibitem[{\citenamefont{Nichol et~al.}(2008)\citenamefont{Nichol, Hemesath,
  Lauhon, and Budakian}}]{nichol2008displacement}
\bibinfo{author}{\bibfnamefont{J.~M.} \bibnamefont{Nichol}},
  \bibinfo{author}{\bibfnamefont{E.~R.} \bibnamefont{Hemesath}},
  \bibinfo{author}{\bibfnamefont{L.~J.} \bibnamefont{Lauhon}},
  \bibnamefont{and} \bibinfo{author}{\bibfnamefont{R.}~\bibnamefont{Budakian}},
  \bibinfo{journal}{Applied Physics Letters} \textbf{\bibinfo{volume}{93}},
  \bibinfo{pages}{193110} (\bibinfo{year}{2008}).

\bibitem[{\citenamefont{Kubo}(1966)}]{Kubo1966}
\bibinfo{author}{\bibfnamefont{R.}~\bibnamefont{Kubo}}, \bibinfo{journal}{Rep.
  Prog. Phys.} \textbf{\bibinfo{volume}{29}}, \bibinfo{pages}{255}
  (\bibinfo{year}{1966}).

\bibitem[{\citenamefont{Caniard et~al.}(2007)\citenamefont{Caniard, Verlot,
  Briant, Cohadon, and Heidmann}}]{caniard2007observation}
\bibinfo{author}{\bibfnamefont{T.}~\bibnamefont{Caniard}},
  \bibinfo{author}{\bibfnamefont{P.}~\bibnamefont{Verlot}},
  \bibinfo{author}{\bibfnamefont{T.}~\bibnamefont{Briant}},
  \bibinfo{author}{\bibfnamefont{P.-F.} \bibnamefont{Cohadon}},
  \bibnamefont{and} \bibinfo{author}{\bibfnamefont{A.}~\bibnamefont{Heidmann}},
  \bibinfo{journal}{Phys. Rev. Lett.} \textbf{\bibinfo{volume}{99}},
  \bibinfo{pages}{110801} (\bibinfo{year}{2007}).

\bibitem[{\citenamefont{Sheard et~al.}(2004)\citenamefont{Sheard, Gray,
  Mow-Lowry, McClelland, and Whitcomb}}]{sheard2004observation}
\bibinfo{author}{\bibfnamefont{B.~S.} \bibnamefont{Sheard}},
  \bibinfo{author}{\bibfnamefont{M.~B.} \bibnamefont{Gray}},
  \bibinfo{author}{\bibfnamefont{C.~M.} \bibnamefont{Mow-Lowry}},
  \bibinfo{author}{\bibfnamefont{D.~E.} \bibnamefont{McClelland}},
  \bibnamefont{and} \bibinfo{author}{\bibfnamefont{S.~E.}
  \bibnamefont{Whitcomb}}, \bibinfo{journal}{Physical Review A}
  \textbf{\bibinfo{volume}{69}}, \bibinfo{pages}{051801}
  (\bibinfo{year}{2004}).

\bibitem[{\citenamefont{Rugar~{\it et al.}}(2004)}]{Rugar2004}
\bibinfo{author}{\bibfnamefont{D.}~\bibnamefont{Rugar~{\it et al.}}},
  \bibinfo{journal}{Nature} \textbf{\bibinfo{volume}{430}},
  \bibinfo{pages}{329} (\bibinfo{year}{2004}).

\bibitem[{\citenamefont{Novotny}(2010)}]{novotny2010strong}
\bibinfo{author}{\bibfnamefont{L.}~\bibnamefont{Novotny}},
  \bibinfo{journal}{American Journal of Physics} \textbf{\bibinfo{volume}{78}},
  \bibinfo{pages}{1199} (\bibinfo{year}{2010}).

\bibitem[{\citenamefont{Bohren and Huffman}(1983)}]{Bohren1983}
\bibinfo{author}{\bibfnamefont{C.~F.} \bibnamefont{Bohren}} \bibnamefont{and}
  \bibinfo{author}{\bibfnamefont{D.}~\bibnamefont{Huffman}},
  \emph{\bibinfo{title}{Absorption and Scattering of Light by Small Particles}}
  (\bibinfo{publisher}{WileyVCH, Berlin}, \bibinfo{year}{1983}).

\bibitem[{\citenamefont{Heinrich et~al.}(2011)\citenamefont{Heinrich, Ludwig,
  Qian, Kubala, and Marquardt}}]{Heinrich2011}
\bibinfo{author}{\bibfnamefont{G.}~\bibnamefont{Heinrich}},
  \bibinfo{author}{\bibfnamefont{M.}~\bibnamefont{Ludwig}},
  \bibinfo{author}{\bibfnamefont{J.}~\bibnamefont{Qian}},
  \bibinfo{author}{\bibfnamefont{B.}~\bibnamefont{Kubala}}, \bibnamefont{and}
  \bibinfo{author}{\bibfnamefont{F.}~\bibnamefont{Marquardt}},
  \bibinfo{journal}{Phys. Rev. Lett.} \textbf{\bibinfo{volume}{107}},
  \bibinfo{pages}{043603} (\bibinfo{year}{2011}).

\bibitem[{\citenamefont{Zhang et~al.}(2012)\citenamefont{Zhang, Wiederhecker,
  Manipatruni, Barnard, McEuen, and Lipson}}]{Zhang2012}
\bibinfo{author}{\bibfnamefont{M.}~\bibnamefont{Zhang}},
  \bibinfo{author}{\bibfnamefont{G.~S.} \bibnamefont{Wiederhecker}},
  \bibinfo{author}{\bibfnamefont{S.}~\bibnamefont{Manipatruni}},
  \bibinfo{author}{\bibfnamefont{A.}~\bibnamefont{Barnard}},
  \bibinfo{author}{\bibfnamefont{P.}~\bibnamefont{McEuen}}, \bibnamefont{and}
  \bibinfo{author}{\bibfnamefont{M.}~\bibnamefont{Lipson}},
  \bibinfo{journal}{Phys. Rev. Lett.} \textbf{\bibinfo{volume}{109}},
  \bibinfo{pages}{233906} (\bibinfo{year}{2012}).

\bibitem[{\citenamefont{Seok et~al.}(2013)\citenamefont{Seok, Buchmann, Wright,
  and Meystre}}]{Seok2013}
\bibinfo{author}{\bibfnamefont{H.}~\bibnamefont{Seok}},
  \bibinfo{author}{\bibfnamefont{L.}~\bibnamefont{Buchmann}},
  \bibinfo{author}{\bibfnamefont{E.~M.} \bibnamefont{Wright}},
  \bibnamefont{and} \bibinfo{author}{\bibfnamefont{P.}~\bibnamefont{Meystre}},
  \bibinfo{journal}{arXiv:1309.7134}  (\bibinfo{year}{2013}).

\bibitem[{\citenamefont{Bauer et~al.}(2013)\citenamefont{Bauer, Orlov, Peschel,
  Banzer, and Leuchs}}]{Bauer2013}
\bibinfo{author}{\bibfnamefont{T.}~\bibnamefont{Bauer}},
  \bibinfo{author}{\bibfnamefont{S.}~\bibnamefont{Orlov}},
  \bibinfo{author}{\bibfnamefont{U.}~\bibnamefont{Peschel}},
  \bibinfo{author}{\bibfnamefont{P.}~\bibnamefont{Banzer}}, \bibnamefont{and}
  \bibinfo{author}{\bibfnamefont{G.}~\bibnamefont{Leuchs}},
  \bibinfo{journal}{Nature Photon.}  (\bibinfo{year}{2013}).

\bibitem[{\citenamefont{Chen et~al.}(2011)\citenamefont{Chen, Ng, Lin, and
  Chan}}]{Chen2011}
\bibinfo{author}{\bibfnamefont{J.}~\bibnamefont{Chen}},
  \bibinfo{author}{\bibfnamefont{J.}~\bibnamefont{Ng}},
  \bibinfo{author}{\bibfnamefont{Z.}~\bibnamefont{Lin}}, \bibnamefont{and}
  \bibinfo{author}{\bibfnamefont{C.}~\bibnamefont{Chan}}, \bibinfo{journal}{Nat
  Photon} \textbf{\bibinfo{volume}{5}}, \bibinfo{pages}{531}
  (\bibinfo{year}{2011}).

\end{thebibliography}
\end{document}